# Modelling Competing Legal Arguments using Bayesian Model Comparison and Averaging

Martin Neil, Norman Fenton, David Lagnado, Richard David Gill

March 5th, 2019

**Abstract**

Bayesian models of legal arguments generally aim to produce a single integrated model, combining each of the legal arguments under consideration. This combined approach implicitly assumes that variables and their relationships can be represented without any contradiction or misalignment, and in a way that makes sense with respect to the competing argument narratives. This paper describes a novel approach to compare and 'average' Bayesian models of legal arguments that have been built independently and with no attempt to make them consistent in terms of variables, causal assumptions or parameterization. The approach involves assessing whether competing models of legal arguments are explained or predict facts uncovered before or during the trial process. Those models that are more heavily disconfirmed by the facts are given lower weight, as model plausibility measures, in the Bayesian model comparison and averaging framework adopted. In this way a plurality of arguments is allowed yet a single judgement based on all arguments is possible and rational.

# 1. Introduction

Typically, Bayesian models of legal arguments have been developed with the aim of producing an integrated model which combines each of the legal arguments under consideration, such as those presented by the defence and prosecution in a trial [1], [2], [3], [4], [5], [6], [7], [8]. This approach implicitly assumes that the resulting integrated model can represent variables and their relationships without any contradiction or misalignment and in a way that makes sense with respect to the competing argument narratives. We call this approach to legal argumentation the "integrated Bayesian perspective" (see [8], for an account of the history and status of this research area). However, the integrated approach can be challenging in practice. By seeking to unify disparate arguments in a single consistent model we encounter modelling difficulties that are hard to overcome, such as those reported in [9] relating to the basic requirement of mutual exclusivity, and the requirement that conditional or causal dependencies remain consistent despite competing or contradictory argument narratives. Finally, the integrated approach assumes an omniscient fact-finder capable of rationally fusing all relevant information all at once when, in practice, the fact-finder is part of an evolving legal process that culminates in a decision.

Whilst the integrated approach represents a noble ideal for determining the 'true' state of the world, we can find no practical requirement or legal stipulation to adopt the integrated approach and neither can we assume that, for any legal case, there are only ever two competing arguments requiring unification. Indeed, each party in a trial process may present more than one argument, each mutually exclusive of the other, positing different causal conjectures, assigning different weights to evidence or even ignoring some kinds of evidence altogether.

Indeed, non-Bayesian approaches to legal argumentation have tended to be narrative-based with a focus on comparisons between competing stories and explanations with much less emphasis on formal integration [2], [3], [4], [9]. Recent work that incorporates scenario-based approaches with Bayesian networks has attempted to partly address this problem [10], [11]. Likewise, a convincing attempt at integrating narrative and probabilistic perspectives has been presented in [12], with an emphasis on modelling more than mere evidence but also considering competing narratives, explanations and notions of credibility and resiliency. However, the main weakness in the approach taken in [12] is that it fails to offer a convincing and operational means to structure and compare competing narratives.

More recent work [13], [14] has attempted to connect arguments, probability and scenarios-based approaches.

This paper presents an approach to modelling legal arguments that maintains the separation of each legal argument in separate Bayesian network models (described in Section 3). This approach allows differences in the variable definitions and causal dependencies that each argument may contain. Thus, in principle, defence and prosecution models may contain different variables with radically different causal dependencies and dissimilar probability assignments. Additionally, the paper aims to be consistent with the hypothetico-deductive method in that more accurate empirical inferences made by one legal argument, rather than another competing legal argument, are given more weight.

The overall objective of the approach proposed is to model legal decision-making from the perspective of an observer or fact-finder (such as a judge or jury member) who observes the different arguments and facts presented by both sides of a case. Such an observer will formulate prior beliefs about the integrity and coherence of the arguments and will then revise their beliefs after they observe witnesses present their evidence and defend it under cross-examination. The observer's subsequent belief in the credibility of the witnesses will drive their revised belief in the credibility of the narratives.

This paper is structured as follows: Section 2 describes and motivates the underlying ideas and Section 3 summarises the Bayesian modelling approaches underpinning our framework. Section 4 presents the proposed framework. In Section 5 we use an example to show how the framework might be applied as a trial develops. Section 6 presents an integrated model and discusses how it compares with the example results. Finally, in Section 7 we offer some discussion and then give conclusions in Section 8.

## 2. The underlying idea and its motivation

The aim of our paper is to compare Bayesian models of legal arguments, even if they have been built independently and with no attempt to make them consistent in terms of variables, causal assumptions or parameterisation. We consider the situation in a criminal trial, where a suspect has been charged with some crime, and where prosecution and defence arguments each can be represented by a Bayesian model. Several facts (primary pieces of evidence), such as the results of forensic analysis and witness statements, have been established; but the weight, relevance and interpretation of those facts are disputed by the two sides. During court hearings, by cross-examination and argument, further evidence of secondary nature comes to light which we call source credibility evidence. This source credibility evidence may or may not change our judgement of source credibility. How reliable is a witness? What kinds of errors can be made when calling on forensic evidence?

We suppose that the Bayesian model produced by each of the two parties allows them to express the arguments about both guilt and innocence of the defendant[1]. Each is a model of the joint probability distribution of several random variables: some of which are supposed to have been observed, others of which ("hidden variables") have not. The variables in the two models need not be the same since the two parties have different pictures of the causality relations between what is observed in the real world on the one hand, and the guilt or innocence of the defendant on the other hand. The model of the prosecution should somehow explain the facts by taking the defendant to be guilty, while the model of the prosecution is an explanation of the same facts by taking the defendant to be innocent. But the two models do not typically assume *a priori* guilt and innocence respectively. They might even contain the same *a priori* probabilities of guilt. Typically, prosecution and defence will both have arguments concerning motive and opportunity, which describe the situation in the real world at times prior to the crime, as well as arguments about forensic evidence, which typically bears on actions or events at the time of the crime. If the model of the prosecution deals with evidence for (i.e., facts relating to) motive and opportunity, then it must admit that absent those facts, innocence must have been a real possibility, hence it will implicitly imply a non-zero prior probability of innocence; by prior we mean prior to knowledge of the facts. The prosecution model is put forward to show that *a posteriori*, i.e., given the facts, innocence is highly unlikely. Conversely, the defence model is put forward to show that *a posteriori*, thus given the same facts, innocence is quite possible.

---
[1] Note that throughout this paper we use argument and model interchangeably, since we propose that each argument can be represented by a model.

Both models are in fact models for the joint probability distribution of guilt/innocence and the facts. They each define their own prior for guilt, as well as their own conditional probabilities for the facts given guilt or innocence. We consider a third party, the fact-finder, who has to evaluate the arguments of the two sides. This could be a judge, or judges' bench, or a jury, or an academic studying historical cases to model either actual or ideal behaviour of triers of fact. Whether our framework should be considered descriptive or prescriptive is left open. As we said, both models should allow guilt and innocence to be expressed within the model, and both models should allow us to express realisation or actualisation of certain facts. The various hidden variables in the two models model the causal relations between what can be observed in a way which is consistent with our understanding of "how the world works".

An argument that fails to explain any of the facts of the case or any supplementary facts that arise during the case should be believed less than an argument that successfully explains all facts. Therefore, arguments that do not explain facts should be penalised by the fact-finder. However, unlike in science where models are judged by how well they predict facts, here legal arguments are partly and initially "fitted" to the facts and thus do not have the status of scientific predictions. Nonetheless, some underlying operating conditions are shared between science and law: in a court case an argument is tested by cross examining witnesses. This process may reveal new, previously unknown, contradictory facts that discredit the sources used in an argument and hence undermine the argument itself. This process closely parallels scientific practice and so our view is to judge legal arguments in a scientific way based on how well they explain (historical) facts and how well they predict (new) facts revealed during the trial. Indeed, it has been cogently argued that the adversarial fact finding process, as practiced in England and Wales, shares the same Enlightenment values and methods of enquiry as the empirical scientific process [15].

We propose that the ***plausibility*** the fact-finder has in a model should be a function of how well it explains the facts of the case and by its ability to anticipate future facts, revealed during the case. We propose measuring plausibility as the joint probability that all facts are confirmed by a model. Under this framework a model that makes incorrect predictions would suffer penalty and a model that relies on too many auxiliary hypotheses and assumptions will be more fragile and easier to refute. There is persuasive evidence to support the idea that lay people think this way too [16].

Contradictions between hypotheses that attempt to explain old facts and new facts are mediated in our framework by the credibility the fact-finder places in the sources of those facts. If a source is discredited by new facts that contradict older ones, then those elements of the argument that rely on that source will be disbelieved. Hence, source credibility and observations (facts) about source credibility play a prominent role. Accordingly, the original Bayesian models produced by each party are enhanced by the fact-finder such that it reflects their own judgements about these factors.

## 3. Bayesian Inference methods applied

There are two types of Bayesian inference in our framework. The first is *Bayesian Model Comparison and Averaging* (BMCA) and the second are *Bayesian Networks* (BNs). Bayes Theorem underpins both approaches and is a sufficiently good starting point to understand BMCA and BNs. Bayes Theorem states that the probability of a hypothesis variable, $H = h_i$, with $i = 1, \dots, n$ states, given evidence variable, $E$, is:

$$P(H = h_i \mid E) = \frac{P(E \mid H = h_i)P(H = h_i)}{\sum_{i=1}^{n} P(E \mid H = h_i)P(H = h_i)} \quad (1)$$

where: $P(H = h_i \mid E)$ is the *posterior* probability of the hypothesis being true; $P(E \mid H = h_i)$ is the likelihood of observing the evidence, $E$, given the hypothesis; $P(H = h_i)$ is the *prior* probability of the hypothesis being true; and $P(E) = \sum_{i=1}^{n} P(E \mid H = h_i)P(H = h_i)$ is the probability of observing the evidence over all hypotheses.

BMCA is widely used to compare and average predictions, sourced from different models that, despite differences in content and accuracy, allow their combination using the same observed data [17], [18]. BMCA applies a standard norm of scientific investigation by identifying those models that predict or explain the available data, more or less well and weighs each model by the plausibility of their performance. Assuming we have several different models, each competing to explain and predict the same phenomena, the decision maker can then either select the 'best' model, or by producing an ensemble model, average all the model predictions, weighted by the performance of each. We can 'solve' the BMCA problem using Bayes Theorem by replacing hypotheses with models, $M = m_i$, and evidence with data, $D$:

$$P(M = m_i \mid D) = \frac{P(D \mid M = m_i)P(M = m_i)}{\sum_{i=1}^{n} P(D \mid M = m_i)P(M = m_i)} \quad (2)$$

where: $P(D \mid M = m_i)$ is the likelihood of observing that data under model; $M = m_i$; and $P(M = m_i)$ is the prior probability of the model being true. Assuming equal priors, those models with higher likelihood probabilities will have equivalently higher posterior probabilities. We can then select and use the most plausible model as being equal to the one with the highest posterior probability. Alternatively, we can average the resulting predictions for a given variable of interest, $\phi$, made by the ensemble of models:

$$P(\phi \mid D) = \sum_{i=1}^{n} P(\phi \mid M = m_i, D)P(M = m_i \mid D) \quad (3)$$

where the posterior probability, $P(\phi \mid D)$, is the average of the predictions sourced from each model and weighted by their performance using formula (2).

A BN (also known as a graphical probabilistic model) is a Bayesian model that is composed of a graphical structure and a set of parameters. The graphical structure of a BN is a Directed Acyclic Graph (DAG). Each node of the DAG represents a random variable and each directed edge represents a relation between those variables. When one or more parent nodes are connected by a directed edge to a child node a set of probability assignments (the parameterization) is used to define their Conditional Probability Distribution (CPD). An example BN is the joint distribution $P(A, B, C, D) = P(D \mid C)P(B \mid A, C)P(A)P(C)$, shown by Figure 1. Bayes theorem can then be applied to the DAG to query any probability from the model (e.g. $P(A, C) = \sum_{B,D} P(D \mid C)P(B \mid A, C)P(A)P(C)$).

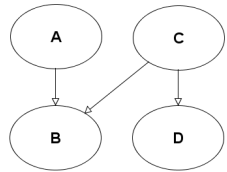

Figure 1 A simple 4 variable BN

The DAG encodes conditional independence assertions between its variables. For example, a node is conditionally independent from the rest of the BN given that its parents, children and the parents of their children are observed (see [19] and [20] for more information on BNs and their conditional independence properties). The conditional independence assertions encoded in the DAG enables a BN to represent a complex joint probability distribution in a compact and factorized way. In some circumstances the DAG may be chosen to represent causal assertions connecting known and unknown events and the CPTs assigned may represent our subjective beliefs (probabilities) about these events. In Figure 1 the factorization $P(D \mid C)P(B \mid A,C)P(A)P(C)$ may therefore be considered as assertion that variables $A, C$ jointly cause variable $B$, while $D$ is a consequence of $C$ alone.

BNs have established inference algorithms that make exact and approximate inference computations by exploiting the conditional independence encoded in the structure. Popular exact algorithms, such as the Junction Tree algorithm [20], and available in commercial and free software packages, [21], provide efficient solutions computations for BNs with only discrete variables.

## 4. Our Framework

Our framework is a direct application of BMCA, stylised to fit a legal context. We assume legal arguments are represented by BN models whose outputs, in the form of plausibility probabilities and assessments of guilt/innocence, are used by BMCA to compare and average. For simplicity we assume there are two sides to the case, each presenting a single defence and a prosecution argument, where each of these can be represented by a single BN model. Yet, in practice more arguments may be 'in play'.

Denoting the two BN models respectively by $m_P$ and $m_D$, we shall take the Bayesian point of view that the fact-finder's uncertainty about which model is correct is expressed by prior probabilities. Note that the prior distribution over models is not the same as a prior distribution of guilt/innocence.

The two sides' arguments each specify a joint probability distribution of a whole list of further variables, conditional on a random variable $M$ taking the value either $m_P$ or $m_D$. Two BN models for the defence and prosecution arguments are shown in Figure 2, for the example we will cover in detail in Section 5.

The two lists of model variables can differ, but they do have some commonalities - after all, they are both models for the same case. We will list those commonalities later, but first we make some remarks concerning notation. Here we adopt the notational convention that random variables are denoted by upper case letters, realised values of random variables are denoted with lower case letters. We take the Bayesian point of view that probabilities which depend on unknown parameters are just conditional probabilities. Hence, mathematically, unknown parameters are just unobserved values of random variables. We denote guilt by a Boolean variable $G$ (where $G = false$ represent innocence); we take the collection of observed facts to be $F = f$. Furthermore, we denote by $C$ the collection of expressions of credibility of each source of observed facts, $F = f$, where each variable in $C$ is simply the belief in the accuracy of a source of evidence. We can differentiate credibility in the two models respectively as $C = C_P$ and $C = C_D$. We can do the same for fact nodes such that we have $F = F_P$ and $F = F_D$ because models may agree on some facts but not on others. Likewise, a fact variable may be agreed by both sides or presented by one side but unchallenged by the other, and hence have no associated credibility variable in one or both models (hence a credibility variable may not appear in one model but do so in another if its associated fact variables are ignored in one but not the other).

Each model will include one or more hypothesis variables, $H = h$, representing unknowns (such as whether some event happened or not), that are causally associated with any observed facts providing evidence about the hypotheses. Each side will share at least one hypothesis – guilt, $G$, but will differ in the number and role of additional hypotheses nodes included in their model. Within our framework we therefore give special status to guilt, $G$, to differentiate between it and other hypothesis variables, $H$. However, despite their necessary role in each model unknown hypothesis variables play no prominent part in our framework and are therefore removed by marginalization.

Both models must include variable $G$ and collections of variables $F, C$ and ranging over the same possible values. But everything else can be different. Prosecution and defence agree on the facts, but they disagree on their meaning and interpretation in numerous ways: they have different understandings of how the facts are causally related to guilt/innocence, both qualitatively (DAGs) and quantitatively (CPTs); different hypothesis variables can be involved; and finally, they have different *a priori* positions concerning the reliability of different sources of evidence, but ultimately it is the fact-finder's assessment of these that is pre-eminent.

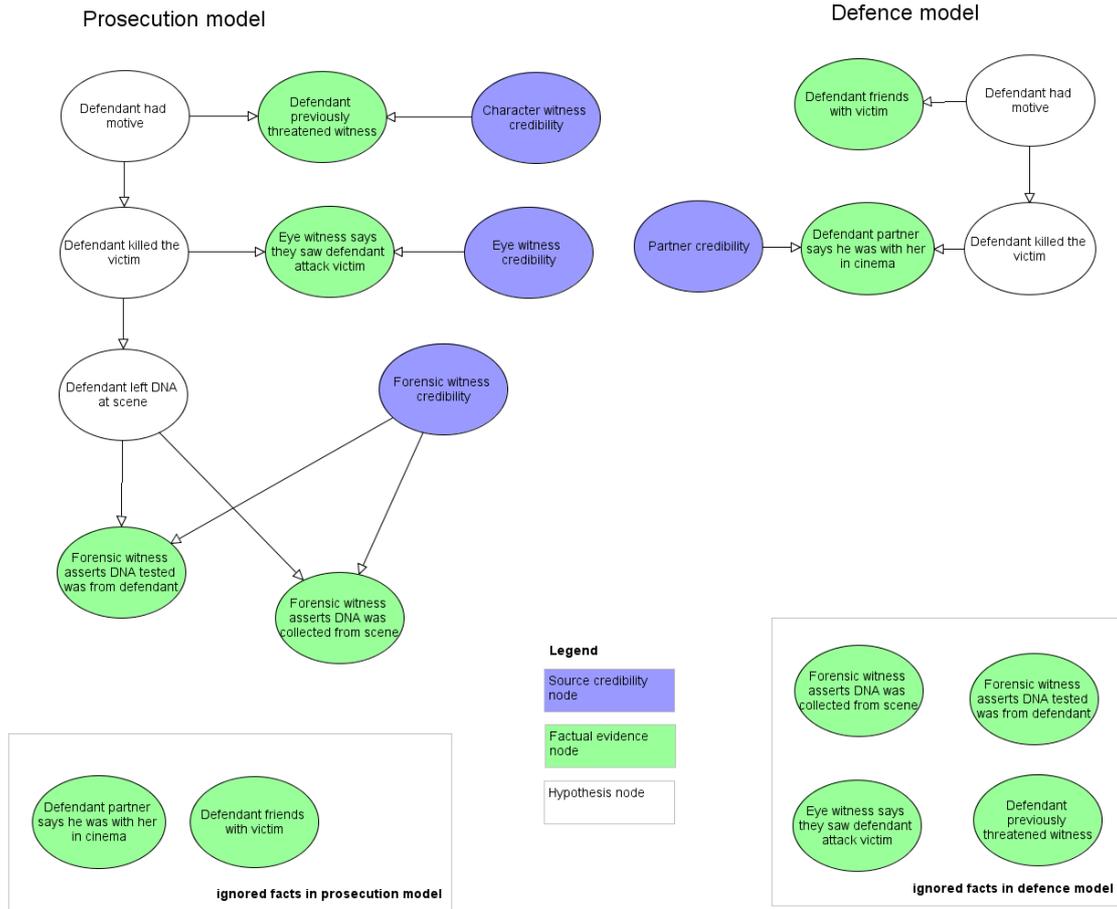

Figure 2 Initial prosecution, $m_P$, and defence models, $m_D$

Marginalizing over the hypothesis nodes, the prosecution's Bayesian model determines a joint probability distribution of $(G, F, C)$ conditional on $M = m_P$. Similarly, the defence's model determines a joint probability distribution of $(G, F, C)$ conditional on $M = m_D$. Together with a prior distribution of M this gives us a joint probability distribution of $(M, G, F, C)$.

The prosecution believes that $C = C_P$ and the defence believes that $C = C_D$. Each believes that their own model is correct, conditioned on the facts and the credibility of the sources of these facts. They then assert that the defendant is guilty or innocent, respectively, because:

(prosecution)    $P(G = guilty \mid F = f, C = C_P, M = m_P)$ is large (4)
(defence)        $P(G = guilty \mid F = f, C = C_D, M = m_D)$ is small (5)

The court hearings allow the fact-finder to get their own opinion as to the value of $C$. We suppose that this opinion is summarised by a definitive understanding that $C = c$. We furthermore define *model plausibility* as the probability of observing the facts given model $m$ is true, the credibility of the sources of those facts, conditioned on the model conclusion made by each party (guilt or innocence):

(prosecution)    $P(F = f \mid C = c, G = guilty, M = m_P)$ (6)

(defence)        $P(F = f \mid C = c, G = \neg guilty, M = m_D)$ (7)

The model plausibility "belongs" to the fact-finder - the party who must make assessments of the likelihood that $F = f$.

From now on we will abuse notation in the conventional way by rewriting model plausibility as:

$$P(F = f \mid C = c, G = g, M = m) = p_{F|C,M}(f \mid c, g, m) = p(f \mid c, g, m) \quad (8)$$

Using the definition of conditional probability, we obtain the following result[2]:

$$p(m \mid c, g, f) = \frac{p(f \mid c, g, m) p(c, g \mid m) p(m)}{\sum_{m'} p(f \mid c, g, m') p(c, g \mid m') p(m')} \quad (9)$$

Which shows how model plausibility, $p(f \mid c, g, f)$, impacts on our posterior belief in which model is true, $p(m \mid c, g, f)$. Moreover, model plausibility also enters our posterior belief in guilt or innocence (writing $g$ for the value 'guilty' of the variable $G$), resulting in the theorem:

$$p(g \mid c, f) = \frac{\sum_m p(g \mid c, f, m) p(m \mid c, g, f)}{\sum_m p(m \mid c, g, f)} = \frac{\sum_m p(g \mid c, f, m) p(f \mid c, g, m) p(c, g \mid m) p(m)}{\sum_m p(f \mid c, g, m) p(c, g \mid m) p(m)} \quad (10)$$

Essentially, the court proceedings might well give the fact-finder more insights into the two models. Can the assumed causal structures be taken seriously, along with all their hidden variables; are the accompanying causal relations acceptable? It seems to us that the court proceedings could lead the fact-finder to wish to adjust the two models or to re-evaluate the prior over the two models. More seriously still, the fact-finder might conclude that, even after such modifications, neither model can be accepted. If the two possibilities are not exhaustive, then a probability of guilt given that just one of the two is true could be terribly misleading; we will return to this issue later in Section 5. Also, note that the prior distribution over the two models is not the same as the prior probability of guilt or innocence. It is rather some kind of meta-prior: the subjective prior probability that one or the other of two whole complexes of arguments is correct. That makes it all the harder to evaluate, and more likely that further analysis could reveal large inadequacies of both models. Significant differences between assumptions about causal structure could make clear that different aspects of both models must be rejected. The scientific method tells us to reject models which make false predictions, but it does not tell us how to amend the models when that happens, and Bayesian methodology is not much help here. These and other open questions are revisited at the end of the paper.

Our framework makes use of the evidence accuracy idiom described in [4], which represents how a fact-finder reasons about evidence reliant on the credibility of a source providing that evidence. It is implemented as a BN to distinguish between the truth of a hypothesis (such as whether some event happened or not) and the source (such as a witness) that provides evidence about the hypothesis (which could be direct, such as asserting that the hypothesis is true, or indirect such as making an assertion which supports the hypothesis). Figure 3 presents the basic BN form of this idiom.

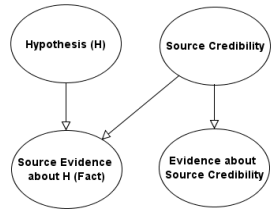

Figure 3 Model showing evidence accuracy idiom

The extent to which the fact-finder can infer the truth of the hypothesis from the source assertion obviously depends on the source credibility. If the source is credible then their evidence will increase the fact-finder's belief in the hypothesis being true; if the source is discredited, then their testimony will have little or no impact on the fact-finder's belief in the hypothesis. In practice, the evidence about source credibility might take the form of oral contradictions, obfuscations, untruths or physical or emotional displays that are taken to betray the credibility of the source; or impeccable credentials and cogent reasoning that make them believable. Evidence about source credibility generally does not fix the values of source credibility nodes. Such evidence (together with the facts $F = f$) only update the prior distributions over their values within each of the prosecution and defence models. This is the sense in which the two sets of variables may be collapsed together; however, it usually will not allow us to instantiate $C$ (with certainty) for both models simultaneously. We can simply consider evidence about source credibility as "just another kind of evidence", i.e., we add it to the facts $F$. The credibility variables $C$ become "hidden variables". We just have two models for the joint distribution of $F$ and $G$, and a prior over the models; both models include many "hidden variables", including the credibility sources. They are used to express the causal relations between what is observed. We compute the posterior probabilities of the models, and the posterior probability of guilt, using even simpler versions of the earlier formulas, for instance:

$$p(g \mid f) = \frac{\sum_m p(g \mid f, m) p(f \mid m) p(m)}{\sum_m p(f \mid m) p(m)} \quad (11)$$

---

[2] For simplicity we use a simple conditioning here, but each model might be represented by different conditioning assumptions. Likewise, to simplify the presentation we have also dropped other variables, such as those representing unknown hypotheses, which are marginalized out in any case.

What has become more complex is that $f$ now stands for the combination of evidence obtained prior to the trial *and* evidence obtained during the trial. If one is interested, one can look at posterior probabilities of source credibility given the evidence $f$. This makes sense given both prosecution and defence models contain the same source credibility nodes in the sense that the probabilities $P(C = c \mid M = m_i)$ are well defined in each model $m$. Indeed, given it is the fact-finder who is judging credibility then it should be the same in each model. If the evidence $f$ consists of two parts $(f_1, f_2)$, one can also use the same formula iteratively, first updating the prior using the information $(F_1 = f_1)$, and then updating the "intermediate posterior" using the information $(F_2 = f_2)$, representing further evidence of a secondary nature coming to light during the case, which we call source credibility evidence. This is, of course, equivalent to using the earlier formula $p(g \mid c, f)$, where the two steps are combined into one.

The defence model might only deal with a subset of the facts. The same can be true for the prosecution. Nor are either side required to challenge the credibility of sources by cross examination. In Bayesian terms this means that a model $m$ is a model for some part $F_m = f_m$ of the facts. We can write $\bar{F}_m$ for the remaining facts, the overline standing for "complement". In this circumstance some variables in a model will be causally disconnected from those others playing a full role in the legal argument. But the defence or prosecution may not just be ignoring $\bar{F}_m = \bar{f}_m$, they may claim irrelevance, which in strict Bayesian terms translates into conditional independence: the defence model might claim independence of the event part $\bar{F}_m = \bar{f}_m$ given innocence and given $F_m = f_m$. The problem is that defining a position on the value of the conditional probabilities for these "ignored facts" is difficult *if they are irrelevant*. As we saw from (7), we do need the values of $p(f \mid m)$ for both models $m$, where $f = (f_m, \bar{f}_m)$ is the combination of all facts, so we cannot simply drop facts from one model but keep them in the other. One solution is to allow the fact-finder to consider the relevance of an ignored fact to a model, either by adjusting the prior belief in the model or by assigning a probability distribution to each ignored fact. If judged genuinely irrelevant, a non-informative distribution might be assigned to each ignored fact, resulting in $p(\bar{f}_m \mid m)$ equaling one half to the power of the number of ignored facts nodes, the dimension of $F_m$. That way, ignoring more facts becomes more heavily penalised. This has the side effect of allowing us to compute the probability of a model making a random guess of the facts, thus providing a baseline. On the other hand, some facts may be ignored for legitimate reasons, and thus may be assigned higher or lower probability, conditional on background knowledge. In these ways ignored facts play a crucial role in our framework.

In law, the defence is not obliged to come up with a complete scenario dealing thoroughly with all facts brought to the case. As we have said the defence's position might well be that the accused is innocent, and certain facts are simply quite irrelevant. Therefore, the fact-finder should be allowed flexibility to deal with what we call "silent facts". Silent facts may represent a deliberate strategy to avoid self-incrimination, or, alternatively, where the defendant simply has not had access to the resources or time needed to muster a credible counter argument to explain these facts. Likewise, there may be suspicion that the prosecution might be suppressing evidence to help secure a conviction. These reasons serve to "explain away" ignored facts in a model which, although they may be missing from the argument made, are nevertheless very informative. We suggest handling this by allowing the fact-finder to construct explanations for ignored facts that reflect their beliefs about the reasons for silence i.e. they explain them away by extending the causal model.

Of course, during a court case, or indeed during the process of investigation, arguments do not remain static. They change as defence and prosecution react to new evidence. At its most basic we can consider two kinds of narratives: the story of the crime (and its investigation) and the story of the court case [22], [23]. In one we have causal conjectures about what happened and in the other we deal with causal conjectures that seek to undermine or support the first. In the latter the timing of the presentation of evidence, as well as the type and strength of evidence, can be crucial in testing the causal narrative about the crime. For instance, an advocate may keep some information back for cross examination in order to unbalance a witness or they may call a witness who presents 'surprising' testimony that may overturn the opposing case [24]. Similarly, at various stages in the legal process one party may be privy to information not available to the other. This information asymmetry is rightly considered unfair in most legal jurisdictions and in such cases all information must be disclosed to all parties. Interestingly, this unfairness property is mimicked in Bayes' because we cannot carry out Bayesian model comparison if the models are being compared against different data. So, in this way Bayes theorem enforces the legal requirement of fairness. It is our intention that our framework be flexible enough to deal with dynamic shifts in the case and asymmetry.

In our framework some probabilities are provided by the fact-finder, as inputs, and some are computed, as outputs, either manually using the mathematics of probability theory or automatically using Bayesian Network software such as [21]. Thus in (9) the fact finder estimates the probabilities $p(f \mid c, g, m)$, $p(c, g \mid m)$ and $p(m)$ and these are then used as inputs, via software or manually, to compute $p(m \mid c, g, f)$ and $p(g \mid f)$.

As a final word it should be clear that, in practice, each legal argument will be constructed independently and with different objectives in mind, and so there is no guarantee that the variables and states specified in one are identical or consistent with those presented in another. An example where this is obvious occurs when a prosecutor may define the variable for guilt, $G$, as having mutually exclusive states $\{murder, \neg murder\}$ whilst the defence might specify the variable innocent, $I$, with mutually exclusive states $\{suicide, murder\}$. Here $\neg murder$ includes the possibility of an accident, a state that cannot even be recognised in the defence model since this model admits only two events and neither of which includes accident. Therefore, whilst it cannot be practically guaranteed that variable definitions are uniform and standardized across arguments, we assume that the fact-finder is able to impose some uniformity at least for the purposes of applying this framework.

## 5. Applying the framework to an example

To illustrate the framework, consider the following hypothetical case:

> *A victim is known to have been murdered. A defendant is accused of the murder. The prosecution argument is based on these facts*

- *the defendant previously threatened to kill the victim, and this was witnessed*
- *an eyewitness who claims to have been at the crime scene and asserts to having seen the defendant kill the victim*

- *a forensic expert witness asserts that DNA collected from the crime scene matches that of the defendant.*

*While the defence is silent on the fact that the defendant previously threatened the victim and has no comment on the eyewitness statement, their argument is based on the claim that the defendant was not at the scene of the crime at the time – a claim supported by the defendant's partner who asserts that she was in a cinema with him at the time of the crime. Also, the defence claim that the victim and the defendant were friends (and hence there was no motive)*

These initial arguments may be those represented in the prosecution and defence opening statements. The BN models representing the fact-finder's understanding of these prosecution and defence arguments is shown in Figure 2, along with a legend showing the different types of nodes used. The CPTs for the example are listed in the Appendix; note that this includes the prior values for the credibility nodes which are never instantiated directly with evidence.

The facts, $f$, of the prosecution model, $m_P$, are:

- "Forensic witness asserts DNA collected from scene" = True
- "Forensic witness asserts DNA tested was from defendant" = True
- "Eye witness says they saw defendant attack victim" = True
- "Witness claims that defendant previously threatened them" = True

The facts, $f$, the defence model, $m_D$, are:

- "Defendant partner says he was with her in cinema" = True
- "Witness claims that defendant friends with victim" = True.

Notice that the defence makes no attempt to explain several facts that support the prosecution model. Also, the prosecution does not include the two defence facts. Hence, both models (implicitly) contain ignored facts. These are ignored facts in our framework. Also, notice that the fact "Defendant friends with victim" = True does not have an associated credibility source variable. This is because this fact is introduced in the defence argument but never addressed nor challenged by the prosecution, hence any judgement about source credibility is unnecessary in either model.

Let's assume the fact-finder decides to weigh the models according to how well they explain the facts of the case, giving higher weight to the prosecution model: $p(m_P) = 0.8, p(m_D) = 0.2$. Let's also assume that the fact-finder assigns their own prior beliefs in the source credibility variables, $C = c$, as given in the Appendix.

The facts observed, $f$, update the fact-finder's source credibility variables to provide new posterior beliefs that then affect the inference of $G$ in each model. By executing the models we compute the marginal probability of guilt directly from the guilt node, $G = guilty\ (True)$, named: "Defendant killed the victim", conditioned on the facts: $p(g\ |c, f, m_P) = 0.999$, $p(g\ |\ c, f, m_D) = 0.0014$. As expected, the fact-finder's perceives that the prosecution argument is very certain of guilt and the defence argument is convinced of innocence.

Under our framework we first measure the plausibility the fact-finder should have in each model. This is the joint probability of all facts given the model, assuming guilt or innocence respectively. For the prosecution the plausibility is the probability of the joint event: {"Forensic witness asserts DNA collected from scene" = True, "Forensic witness asserts DNA tested was from defendant" = True, "Eye witness says they saw defendant attack victim" = True, "Defendant previously threatened witness" = True, "Defendant partner says he was with her in cinema" = True, "Defendant friends with victim" = True} conditioned on $G = guilty$. For the defence it is the same joint event conditioned on , $G = \neg guilty$. Here we assume the fact-finder is happy to assign non-informative distributions to the ignored facts. Putting these observations into our model and equations (6) and (7) gives us:

$$\text{(prosecution)} \quad P(F = f \mid C = c, G = guilty, M = m_P) = 0.330$$

$$\text{(defence)} \quad P(F = f \mid C = c, G = \neg guilty, M = m_D) = 0.050$$

Therefore, the probability of the prosecution argument explaining the facts is 0.33 and for the defence is 0.05. A random assignment of truth values to the facts would yield $P(F = f) = 1/2^6 = 0.015625$, so the prosecution model is significantly better than a guess, but the defence model less so. The plausibility of the defence model is low because of the number of ignored facts in the model: had more facts been explained the plausibility would have been higher.

We now put the model priors and the plausibility probabilities into Equation (6), yielding these posterior beliefs in each model:

$$p(m_P \mid c, g, f) = \frac{p(f \mid c, g, m_P)p(c, g \mid m_P)p(m_P)}{p(f \mid c, g, m_P)p(c, g \mid m_P)p(m_P) + p(f \mid c, g, m_D)p(c, g \mid m_D)p(m_D)}$$

$$= \frac{0.330(0.8)}{0.050(0.2) + 0.330(0.8)} = 0.964$$

$$p(m_D \mid c, g, f) = \frac{p(f \mid c, g, m_D)p(c, g \mid m_D)p(m_D)}{p(f \mid c, g, m_P)p(c, g \mid m_P)p(m_P) + p(f \mid c, g, m_D)p(c, g \mid m_D)p(m_D)}$$

$$= \frac{0.050(0.2)}{0.050(0.2) + 0.330(0.8)} = 0.036$$

So, at the opening of the trial the fact-finder already believes the prosecution model is better at explaining the facts of the case. Next, we need to calculate the probability of guilt given the two models, using equation (10):

$$p(g \mid f) = \frac{\sum_m p(g \mid f, m)p(m \mid f)}{\sum_m p(m \mid f)} = \frac{p(g \mid f, m_D)p(m_D \mid f) + p(g \mid f, m_P)p(m_P \mid f)}{p(m_D \mid f) + p(m_P \mid f)}$$

$$= \frac{0.0367(0.0014) + 0.964(0.999)}{0.0367 + 0.964} = 0.962$$

For the fact-finder, the probability of guilt is 0.962 based on their plausibility in the two models, the credibility of the sources of evidence, the facts presented, and the causal and probabilistic assumptions made in each model.

Our framework has successfully combined two models with different assumptions and produced a single assessment, belonging to the fact-finder, in a way that gives greater weight to the model that explains the facts better. Of course, we also wish to model the dynamic nature of an evolving case, especially using new facts gained during the cross-examination process. As a last step let's now assume that a cross examination has taken place and new supplementary facts have been discovered:

> *The defendant is much more likely to have left DNA at the scene than the prosecution assumes because the defendant was a frequent visitor to the scene of the crime in a way that is consistent with the DNA findings.*

> *During the defence witness cross examination, she claims her partner would have shown up on the cinema's CCTV system. It turns out the police had collected the CCTV video but not made it available to the defence. Subsequently it was made available to the court, and despite not being of high quality, the fact-finder believed it showed someone matching the accused description at the cinema at the time of the crime.*

> *It is revealed by the eye witness that she failed to pick out the defendant on an identity parade*

> *Finally, under cross examination the character witness admitted to being in a rival gang.*

The new *source credibility evidence* (facts), $f_2$, to add to the initial facts in the models, $f_1$, are:

- "CCTV from camera corroborates description" = True
- "Identity parade failure" = True
- "Character witness in rival gang" = True

These new facts lead to newly revised models as shown in Figure 4[3]. In the defence model the facts from the forensic witness do not imply guilt and the CPT for the variable "Defendant left DNA at the scene" assumes it is just as likely to be DNA at the scene if the defendant murdered the victim or not.

Crucially, two facts are now no longer ignored in the defence model: "Forensic witness asserts DNA collected from scene" = True and "Forensic witness asserts DNA tested was from defendant" = True. Likewise, in the prosecution model the evidence "CCTV from camera corroborates description" = True is judged by the fact-finder to be so remote, within the context of the prosecution scenario, that it is assigned a probability value of 0.001: perhaps not unreasonable, given the evidence was supressed.

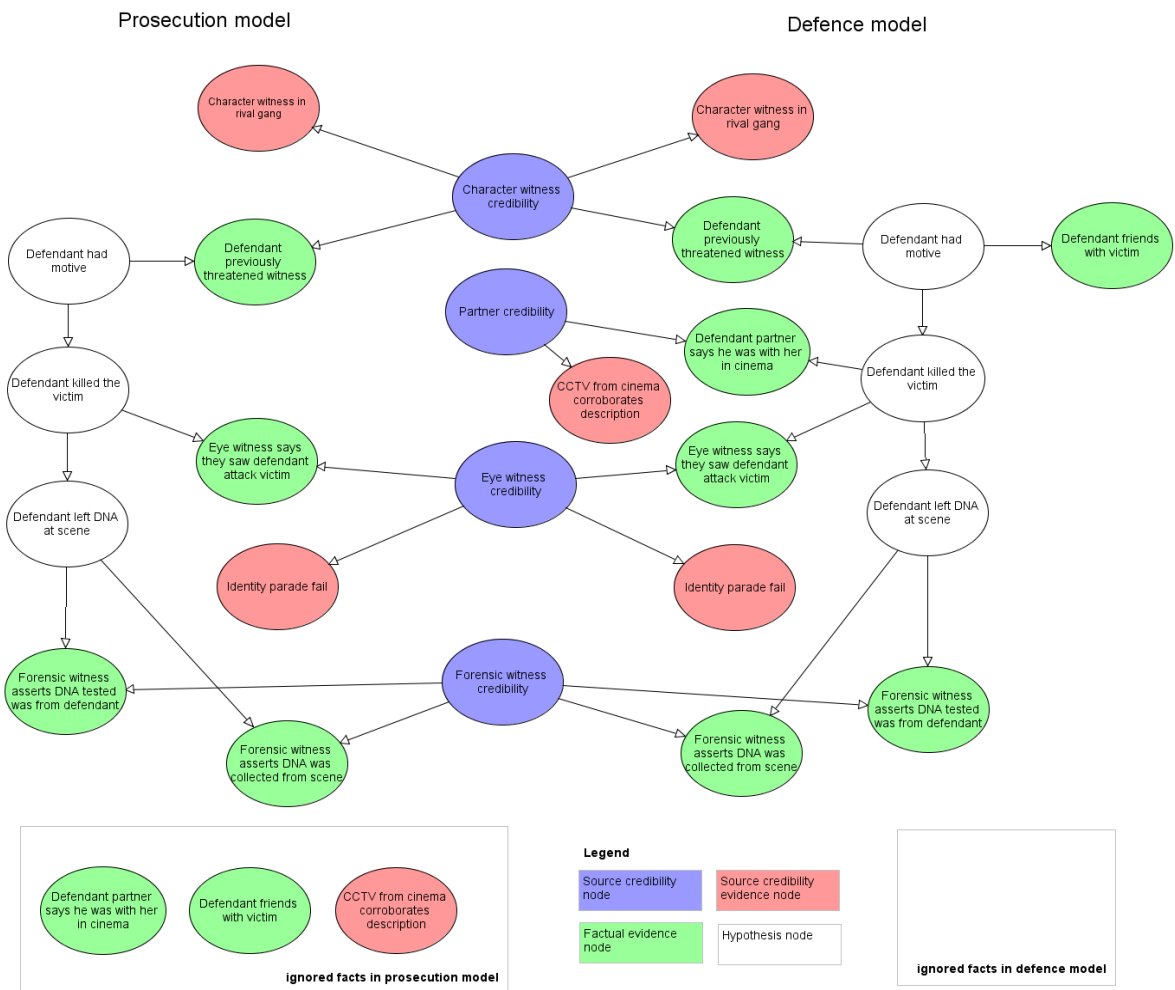

Figure 4 Revised prosecution, $m_P$, and defence models, $m_D$

---

[3] Note that Figure 4 shows two separate BN models, but the underlying computation uses one model. This is because we need to ensure the probabilities computed for the source credibility variables are identical and consistent in each model. So, during computation we combine credibility variables to enable us to produce a single model containing two sub-models linked to a single common collection of credibility nodes belonging to the fact-finder. This does not mean to say that each model contains the same credibility variables; in Figure 4 "Partner credibility" is linked to variables in the defence model only.

In both models, the probability of guilt, conditioned on the facts and fact-finder's beliefs about source credibility, has not changed significantly: $p(g \mid c, f, m_P) = 0.96362, p(g \mid c, f, m_D) = 0.000271$.

We can now update the relevant probabilities for equation (6) and (7) to give:

(prosecution) $\quad P(F = f \mid C = c, G = guilty, M = m_P) = 0.00001$

(defence) $\quad\quad\; P(F = f \mid C = c, G = \neg guilty, M = m_D) = 0.091$

We can now see that the fact-finder's plausibility in the prosecution model has collapsed from 0.33 to 0.00001. Their plausibility in the defence model has increased by a factor of two, from 0.05 to 0.091.

Equations (9) and (10) now give:

$$p(m_P \mid c, g, f) = \frac{p(f \mid c, g, m_P) p(c, g \mid m_P) p(m_P)}{p(f \mid c, g, m_P) p(c, g \mid m_P) p(m_P) + p(f \mid c, g, m_D) p(c, g \mid m_D) p(m_D)}$$
$$= \frac{0.00001(0.8)}{0.091(0.2) + 0.00001(0.8)} = 4.38E - 5$$

$$p(m_D \mid c, g, f) = \frac{p(f \mid c, g, m_D) p(m_D)}{p(f \mid c, g, m_P) p(m_P) + p(f \mid c, g, m_D) p(m_D)} = \frac{0.091(0.2)}{0.091(0.2) + 0.00001(0.8)}$$
$$= 0.999956$$

Thus, the posterior belief in the defence model has risen dramatically from 0.036 to 0.999956 and the posterior belief in the prosecution model has decreased to $4.38E - 5$ from 0.964.

Also, the probability of guilt for the defence model has changed slightly to $p(g \mid c, f, m_D) = 0.0027$ given the new causal structure in the model. The new probability of guilt using equation (10) is:

$$p(g \mid f) = \frac{\sum_m p(g \mid f, m) p(m \mid f)}{\sum_m p(m \mid f)} = \frac{p(g \mid f, m_D) p(m_D \mid f) + p(g \mid f, m_P) p(m_P \mid f)}{p(m_D \mid f) + p(m_P \mid f)}$$
$$= \frac{0.000271 \, (0.999956) + 0.96362(4.38E - 5)}{0.999956 + 4.38E - 5} = 0.000313$$

So, by now explaining the DNA facts and providing damning evidence that cannot be explained by the prosecution, the fact-finder's revised conclusion would dramatically change from $p(g \mid f) = 0.962$ to $p(g \mid f) = 0.000313$. Note this would be a change in the fact-finder's belief rather than that of the advocates.

## 6. Comparison with an integrated Bayesian model

Here we present a Bayesian integrated model developed from our example, using all the information available to the fact-finder up to and including the final step in our example. This integrated model is shown in Figure 5.

Combining the prosecution and defence arguments requires the fact-finder to merge the models using established methods [4], [6]. Bayesian model integration involves the production of a single model including all relevant variables (hypotheses, facts and credibility sources etc) associated with the defence and prosecution positions. By seeking to unify disparate arguments in a single integrated model encounters modelling difficulties that are hard to overcome, such as those reported in [9] relating to the basic requirement of mutual exclusivity, and the requirement that conditional or causal dependencies remain consistent despite competing or contradictory argument narratives. Likewise, the integrated approach assumes an omniscient fact-finder capable of rationally fusing all relevant information. Whilst the integrated approach represents a noble ideal for determining the 'true' state of the world, each party in a trial process may present more than one argument, each mutually exclusive of the other, positing different causal conjectures, assigning different weights to evidence or even ignoring some kinds of evidence or elements of the argument altogether. This last strategy – ignoring some kinds of evidence or argument altogether – does of course reflect a real choice open to the fact-finder and one that has the benefit of simplicity but obviously at the expense of discarding information that the fact finder either doesn't agree with or cannot easily reconcile.

Given that some irresolvable differences may not be easy to ignore then the way to handle these during Bayesian integration is to explicitly model them. This can be done by conditioning subsets of the integrated model representing these irrevocable differences on new scenario variables, representing these differences as uncertain variables. These scenario variables are in turn themselves conditioned on a model variable which directly represents the fact-finder's prior belief in each model. In this way scenario and model variables are interwoven into the BN to connect facts, hypotheses and credibility sources together with the different variables and dependencies associated with the defence and prosecution arguments. This strategy is investigated in [11], [25] but without emphasis on model integration.

In our example three distinct irresolvable differences, or contradictions, arise from the models that are important, and it is worth focusing on how these are handled by an integrated model:

- The defence argument makes distinctly different assumptions about the prior motive of the defendant from that made by the prosecution.
- The assumptions made about the hypothesis variable "DNA left at the scene" differ
- The treatment of 'ignored facts' differs in each model

These differences are reflected in the different CPT tables in each model, as listed in the Appendix. They therefore have an impact on the structure of the integrated model – the integrated model now must accommodate distinct and mutually exclusive sets of assumptions and these are defence or prosecution model dependant. This dependency is represented by explicitly including several conditioning scenario variables that act as "switches" to switch prosecution or defence scenarios on or off. In Figure 5 these scenario variables are shown as the rectangular nodes "Model scene assumption", to represent different assumptions about the past frequency of the defendant visiting the scene, and "Model motive assumption", to represent different assumptions for motive. The third issue relating to how 'ignored facts' are treated also has a significant structural effect on the model, because, again, for each argument the CPTs and conditioning changes. This is accommodated by the scenario variables "Model ignored facts" in the integrated model. Each of these scenario switch variables is then ultimately conditionally dependent on a "Models" variable with mutually exclusive states $M = \{m_P, m_D\}$ (this node replaces the "meta prior" used in our framework).

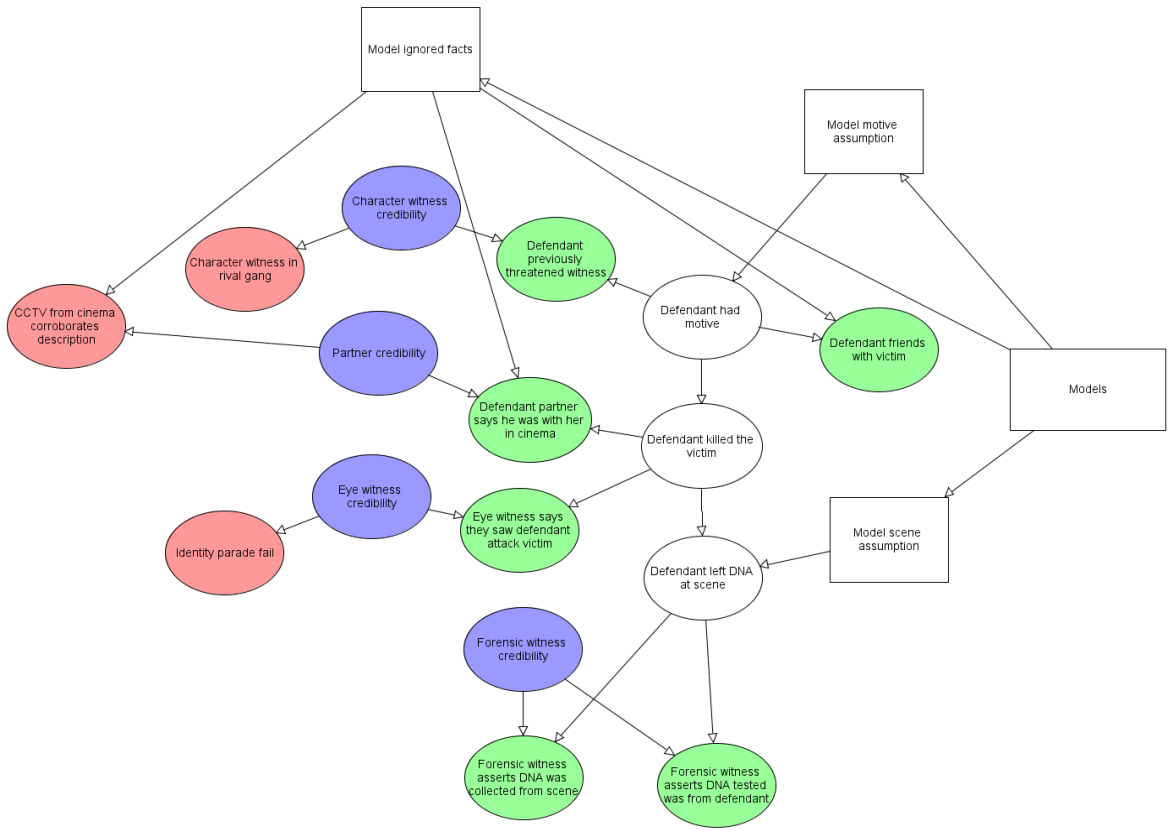

Figure 5 Integrated Bayesian model (model and scenario variables are shown as rectangles)

Obviously, armed with the integrated model we can answer how likely the facts are given the model, and we can infer the belief in the model given the facts and the belief in guilt. Assuming the same prior as before, $p(M = m_P) = 0.8, p(M = m_D) = 0.2$, from the integrated model this is simply calculated to give:

$$p(M = m_P \mid c, g, f) = 0.002397$$

$$p(M = m_D \mid c, g, f) = 0.997603$$

With the result for $G = guilty$:

$$p(g \mid c, f, M) = 0.002849$$

This result looks directly close and comparable to the result calculated using our framework, giving us some comfort that by using different frameworks we can reach similar conclusions.

However, we have argued that it is worthwhile to keep models separate for several reasons: our framework can tolerate differences in causal structure, disagreements of parameterisation and also a difference in the prior beliefs in the arguments. In contrast, from this integrationist example we can see that whilst integration can be enabled using scenario nodes (thus treating a single model as a mixture of different models) this comes at a cost in model legibility and malleability. Also, we would need to add scenario nodes for differences in causal structure and it is a considerable challenge to do this, especially so as the model grows in size and the accompanying potentiality for contradiction.

## 7. Discussion

The framework we have described assumes that exactly one of the two models put forward by prosecution and defence is true. However, in criminal law this may be unrealistic as the fact-finder may come to entertain other pictures of how the world works beyond those initially put forward by defence and prosecution. The fact-finder may arrive at their own picture of how facts in the case are related, and it may well combine some elements of both parties' explanations of the facts, and at the same time reject other elements of both parties' explanations. In other words: the fact-finder is free to come up with an integrated model which, in effect, contradicts both models initially brought to the court by the two parties. In some legal jurisdictions this 'freewheeling' approach may present a problem in that the fact-finder may arrive at conclusions independently of the evidence, arguments and cross examinations related to the case. Indeed, given the way the jury system operates in Anglo-Saxon countries this can and does occur.

A second weakness is that the approach gives little guidance on how to choose the prior distribution over the two models. Because the approach is an unashamed subjective Bayesian approach, the initial probabilities of the two models are subjective or personal probabilities. They are the personal probabilities of the fact-finder. They represent the fact-finder's prior beliefs about the integrity and coherence of the arguments which have been presented by the two parties: they "reflect judgements the fact-finder might make about the global rationality of the argument". Crucially, they do not represent the fact-finder's prior beliefs in guilt or innocence. However, more than just global rationality needs to be evaluated. How is a fact-finder to quantify their prior relative degree of plausibility in the two pictures of the world provided by the two parties in the case? These two pictures are actually very detailed; they consist of more than just an attempt to express rational knowledge about what depends on what in graphical form. They also entail strengths of dependence in precise quantitative form. An outline sketch is transformed into an oil-painting. Also, any remaining uncertainties are quantified and expressed in terms of probability distributions representing, hopefully, rational degrees of belief in different possible values.

Moreover, trial proceedings could well lead to dissatisfaction with both models, even if the prosecution model was significantly more plausible than the defence model. This makes computation of posterior probabilities conditional on just one of the two quite meaningless. In our framework we allow the fact-finder to revise the prior distribution of the two models, as well as their parameters, but there is not a formal (Bayesian or other) way to do this. If the fact-finder truly is trying to identify the true facts of the matter, flexibility and creativity is required. Even if the fact-finder is merely an adjudicator between two fixed points of view, Bayesian thinking cannot tell the fact-finder how to weigh two "wrong arguments". It seems to us that many miscarriages of justice, both in jurisdictions in the adversarial tradition and those in the inquisitorial tradition, have been caused by uncritical acceptance of badly flawed models, even when the defects of those models were explicitly brought to the attention of the court. Subjective confidence in expert evidence can easily depend more on the showmanship of the expert and the simplicity of the expert's message, than on the actual content of the expert's evidence. Similarly, the subjective prior probability of a model could be influenced by the model's simplicity even though it contains logical inconsistencies. We hope our framework might help in analysing such cases.

We recommend our approach as a basis for investigating issues in the comparison of incomparable Bayesian models of legal incompatible arguments; this is, in essence, the problem facing a fact-finder in a criminal case. We do not claim to provide a fool-proof solution. No one does. In the inquisitorial approach the fact-finder is a truth seeker and may creatively generate new models. In the adversarial approach the fact-finder is a referee and is under no obligation to generate new models. One option would be to extend our framework to add a third model, a kind of default model, where the fact-finder should allocate some prior probability to a model in which the accused is "not proven guilty or innocent" (the Scottish model whereby jurors in Scotland can return one of three verdicts: one of conviction, "guilty", and two of acquittal, "not proven" and "not guilty" ). Inconsistencies in the prosecution and defence models could lead to an increase in the prior probability of this third model. Court hearings can reveal inconsistencies in the argument of the prosecution so large, that the prior probability of the default innocence model should increase. Whether or not the defence arguments are reasonable should then become irrelevant. The third, default model, takes over. The subjective Bayesian prior over the models could then only be decided after the court proceedings, not in advance; especially if we allow the parties to modify their models as the trial proceeds.

Clearly what we have proposed is a theoretical framework; while the working example we provided demonstrates that it can be applied in a non-trivial case, we accept that this is very far from being any kind of serious validation of its practicality or usefulness. We hope that such validation will be the subject of future research.

## 8. Conclusions

In previous approaches Bayesian models of legal arguments have been developed with the aim of producing a single integrated model, combining each of the legal arguments under consideration. This combined approach implicitly assumes that variables and their relationships can be represented without any contradiction or misalignment and in a way that makes sense with respect to the competing argument narratives. Rather than aim to integrate arguments into a single model, this paper has described a novel approach to compare and 'average' Bayesian models of legal arguments that have been built independently and with no attempt to make them consistent in terms of variables, causal assumptions or parameterization.

In our framework competing models of legal arguments are assessed by the extent to which the facts reported are confirmed or disconfirmed in court, as judged by the fact-finder. Those models that are more heavily disconfirmed are assigned lower weights, as model plausibility measures, in the Bayesian model comparison and averaging approach adopted. We have presented a simple example to describe the ideas and method and contrasted it with an equivalent integrated Bayesian model.

We believe that our framework approach borrows strengths from the Bayesian and non-Bayesian narrative approaches to legal argumentation without introducing any new significant weaknesses. We would suggest that our approach might be more consistent with legal practice, where plurality in arguments is crucial, yet it does so in a novel way that views elements of the legal process as one consistent with empirical scientific methodology.

## Acknowledgements

This work was supported in part by the ERC (ERC-2013-AdG339182-BAYES_KNOWLEDGE) and the Leverhulme Trust under Grant RPG-2016-118 CAUSAL-DYNAMICS; EPSRC under grant no EP/K032208/1 and the Simons Foundation.

# Appendix

**CPTs for Prosecution BN model at initial stage**

*Defendant had motive, Defendant partner says he was with her in cinema* (ignored fact)*, Defendant friends with victim* (ignored fact)

| | |
|---|---|
| False | 0.5 |
| True | 0.5 |

*Character witness credibility, Eye witness credibility, Forensic witness credibility*

| | |
|---|---|
| False | 0.1 |
| True | 0.9 |

*Defendant previously threatened witness | Character witness credibility, Defendant had motive*

| Character witness credibility | False | | True | |
|---|---|---|---|---|
| Defendant had motive | False | True | False | True |
| False | 0.5 | 0.5 | 0.99 | 0.2 |
| True | 0.5 | 0.5 | 0.01 | 0.8 |

*Defendant killed the victim | Defendant had motive*

| Defendant had motive | False | True |
|---|---|---|
| False | 0.99 | 0.3 |
| True | 0.01 | 0.7 |

*Eye witness says they saw defendant attack victim | Defendant killed the victim, Eye witness credibility*

| Eye witness credibility | False | | True | |
|---|---|---|---|---|
| Defendant killed the victim | False | True | False | True |
| False | 0.5 | 0.5 | 0.99 | 0.01 |
| True | 0.5 | 0.5 | 0.01 | 0.99 |

*Defendant left DNA at scene | Defendant killed the victim*

| Defendant killed the victim | False | True |
|---|---|---|
| False | 0.999999 | 0.0 |
| True | 1.0E-6 | 1.0 |

*Forensic witness asserts DNA tested was from defendant | Forensic witness credibility, Defendant left DNA at scene*

| Forensic witness credibility | False | | True | |
|---|---|---|---|---|
| Defendant left DNA at scene | False | True | False | True |
| False | 0.5 | 0.5 | 0.9999 | 0.0 |
| True | 0.5 | 0.5 | 1.0E-4 | 1.0 |

*Forensic witness asserts DNA was collected from scene | Forensic witness credibility, Defendant left DNA at scene*

| Defendant left DNA at scene | False | | True | |
|---|---|---|---|---|
| Forensic witness credibility | False | True | False | True |
| False | 0.5 | 1.0 | 0.5 | 0.0 |
| True | 0.5 | 0.0 | 0.5 | 1.0 |

**CPTs for Defence BN model at initial stage**

*Defendant had motive*

| | |
|---|---|
| False | 0.99 |
| True | 0.01 |

*Defendant friends with victim | Defendant had motive*

| Defendant had motive | False | True |
|---|---|---|
| False | 0.1 | 0.8 |
| True | 0.9 | 0.2 |

*Partner credibility*

| | |
|---|---|
| False | 0.2 |
| True | 0.8 |

*Defendant killed the victim | Defendant had motive*

| Defendant had motive | False | True |
|---|---|---|
| False | 0.99 | 0.3 |
| True | 0.01 | 0.7 |

*Defendant partner says he was with her in cinema | Partner credibility, Defendant killed the victim*

| Defendant killed the victim | False | | True | |
|---|---|---|---|---|
| Partner credibility | False | True | False | True |
| False | 0.5 | 1.0E-5 | 0.5 | 0.99 |
| True | 0.5 | 0.99999 | 0.5 | 0.01 |

*Forensic witness asserts DNA tested was from defendant* (ignored fact), *Forensic witness asserts DNA was collected from scene* (ignored fact), *Eye witness says they saw defendant attack victim* (ignored fact), *Defendant previously threatened witness* (ignored fact)

| | |
|---|---|
| False | 0.5 |
| True | 0.5 |

**New CPTs for Prosecution BN model at revision stage**

*CCTV from cinema corroborates description* (ignored fact)

| | |
|---|---|
| False | 0.999 |
| True | 0.001 |

*Character witness in rival gang | Character witness credibility*

| Character witness credibility | False | True |
|---|---|---|
| False | 0.01 | 0.99 |
| True | 0.99 | 0.01 |

*Identity parade fail | Eye witness credibility*

| Eye witness credibility | False | True |
|---|---|---|
| False | 0.01 | 0.99 |
| True | 0.99 | 0.01 |

**New CPTs for Defence BN model at revision stage**

*Character witness in rival gang | Character witness credibility*

| Character witness credibility | False | True |
|---|---|---|
| False | 0.01 | 0.99 |
| True | 0.99 | 0.01 |

*Identity parade fail | Eye witness credibility*

| Eye witness credibility | False | True |
|---|---|---|
| False | 0.01 | 0.99 |
| True | 0.99 | 0.01 |

*CCTV from cinema corroborates description | Partner credibility*

| Partner credibility | False | True |
|---|---|---|
| False | 0.999 | 0.001 |
| True | 0.001 | 0.999 |

*Defendant left DNA at scene | Defendant killed the victim*

| Defendant killed the victim | False | True |
|---|---|---|
| False | 0.5 | 0.0 |
| True | 0.5 | 1.0 |

*Forensic witness asserts DNA tested was from defendant | Forensic witness credibility, Defendant left DNA at scene*

| Forensic witness credibility | False | | True | |
|---|---|---|---|---|
| Defendant left DNA at scene | False | True | False | True |
| False | 0.5 | 0.5 | 0.9999 | 0.0 |
| True | 0.5 | 0.5 | 1.0E-4 | 1.0 |

*Forensic witness asserts DNA was collected from scene | Forensic witness credibility, Defendant left DNA at scene*

| Defendant left DNA at scene | False | | True | |
|---|---|---|---|---|
| Forensic witness credibility | False | True | False | True |
| False | 0.5 | 1.0 | 0.5 | 0.0 |
| True | 0.5 | 0.0 | 0.5 | 1.0 |